\documentclass[runningheads,a4paper]{llncs}
\usepackage{url}
\usepackage{subfig}
\usepackage{epsfig}
\usepackage{multirow}
\usepackage{graphics}
\usepackage{graphicx}
\usepackage{algorithmic}
\usepackage{algorithm}
\usepackage{amsfonts}
\usepackage{amsmath}
\usepackage[parfill]{parskip}
\usepackage[applemac]{inputenc}

\usepackage{color}
\usepackage{array}
\usepackage{colortbl}

\newcommand{\ket}[1]{|#1\rangle}

\let\oldsqrt\sqrt
\def\sqrt{\mathpalette\DHLhksqrt}
\def\DHLhksqrt#1#2{%
\setbox0=\hbox{$#1\oldsqrt{#2\,}$}\dimen0=\ht0
\advance\dimen0-0.2\ht0
\setbox2=\hbox{\vrule height\ht0 depth -\dimen0}%
{\box0\lower0.4pt\box2}}

\begin{document}

\mainmatter  % start of an individual contribution

\title{Tree Search and Quantum Computation}
\titlerunning{Tree Search and Quantum Computation}

\author{Luís Tarrataca\inst{1}\thanks{Luís Tarrataca was supported by FCT (INESC-ID multiannual funding) through the PIDDAC Program funds and FCT grant DFRH - SFRH/BD/61846/2009.} \and Andreas Wichert\inst{1}}
\authorrunning{L. Tarrataca \and A. Wichert}

\institute{
GAIPS/INESC-ID\\
Department of Informatics\\
IST - Technical University of Lisbon - Portugal\\
\email{\{luis.tarrataca,andreas.wichert\}@ist.utl.pt}\\
}

%\date{\today}

\maketitle

\begin{abstract}

Traditional tree search algorithms supply a blueprint for modeling problem solving behaviour. A diverse spectrum of problems can be formulated in terms of tree search. Quantum computation, in particular Grover's algorithm, has aroused a great deal of interest since it allows for a quadratic speedup to be obtained in search procedures. In this work we consider the impact of incorporating classical search concepts alongside Grover's algorithm into a hybrid quantum search system. Some of the crucial points examined include: (1) the reverberations of contemplating the use of non-constant branching factors; (2) determining the consequences of incorporating an heuristic perspective into a quantum tree search model.


\end{abstract}

\newcounter{UnitaryOperatorCounter}
\setcounter{UnitaryOperatorCounter}{1}

\section{Introduction \label{sec:introduction}}

The concepts of knowledge representation and the reasoning processes that support knowledge application have long been a key interest area in the field of artificial intelligence. Knowledge enables problem-solving agents to determine an appropriate action in order to better deal with complex environments. Reasoning allows an agent to perform complex decisions by employing a finite amount of knowledge \cite{luger1993}. One such form of reasoning consists of classical tree searching.  Tree search is employed whenever decisions must be made that are based on complex knowledge. Tree search algorithms play a crucial role in many applications, e.g. production systems \cite{post1943} \cite{newell1959} \cite{newell1963} \cite{ernst1969} \cite{anderson1983}  \cite{laird1986} \cite{laird1987}, game playing programs \cite{feldmann1993} \cite{hsu1999} \cite{hsu2002} \cite{campbell2002} and robot control systems \cite{coelho2009a} \cite{coelho2009b}.

The next few section are organized as follows: Section \ref{sec:classicalTreeSearch} reviews some of the major tree search algorithms;  Section \ref{sec:quantumSearch} present an alternative search method based on quantum computation; Section \ref{sec:hybridSystem} proposed an hybrid search system combining classical tree search algorithms alongside quantum search. Section \ref{sec:objectivesAndProblems} presents the objectives and associated problems.

\subsection{Classical tree search \label{sec:classicalTreeSearch}}

The simplest tree search algorithms rely on a ``brute-force'' approach. These methods perform an exhaustive examination of all possible sequences of moves until goal states are reached. The search through the state space systematically checks if the current state is a goal state. If a non-goal state is discovered then the current state is expanded by applying a successor function, generating a new set of states. The choice of which state to expand is determined by a search strategy. In a great deal of occasions, an artificial intelligence application does not possess an adequate level of knowledge enabling the choice of the most promising state. Strategies that can only distinguish between between goal states and non-goal states, without being able to determine if one state is more promising than another, are referred to as uninformed search strategies. Examples of uninformed search strategies include the well known breadth first search \cite{moore1959}, depth-first search \cite{hopcroft1973}, and also the iterative deepening search \cite{slate1977}. Uninformed strategies are only successful for small problem instances. Typically, most problems search space is characterized by an exponential growth \cite{garey1979}. Due to the mammoth dimensions of the search space it becomes impractical, both time- and space-wise, to perform an exhaustive examination.

Alternatively, it is possible to employ additional insights, that arise beyond the definition of the problem. The use of this information, thus the term informed search strategies, allows for solutions to be found more efficiently. Typically, informed search strategies employ an evaluation function $f(n)$ which considers a cost function $g(n)$ alongside a heuristic function $h(n)$. Function $g(n)$ can be interpreted as representing the cost to reach node $n$ whilst $h(n)$ represents an estimate on the cost to reach a leaf node from node $n$. Traditionally, the node with the lowest evaluation value is selected for expansion. Examples of some of the best known informed search strategies include greedy search \cite{newell1965} and A$^{*}$ search \cite{hart1968}. More recent advances on informed strategies include IDA$^{*}$ \cite{korf1985} and RBFS \cite{korf1991}, \cite{korf1993}. A time complexity comparative assessment between the various algorithms is presented in Table \ref{table:treeSearchAlgorithmComparison}.

\begin{table}
\begin{center}
\scriptsize{
\begin{tabular}{| l | c | c | l | } \hline

  \textbf{Search} & \textbf{Reference} & \textbf{Strategy} & \textbf{Time} \\ \hline 
  
  Breadth-first     & \cite{moore1959} 	& Uninformed	& $O(b^{d + 1})$  \\ \hline
  Depth-first 	  & \cite{hopcroft1973} 	& Uninformed   & $O(b^{m})$  	    \\ \hline
  Iterative-deepening &  \cite{slate1977} 	& Uninformed   & $O(b^{d})$ 	    \\ \hline
  Greedy 		  &  \cite{newell1965} 	& Informed	& $O(b^{m})$ 	    \\ \hline
  A$^{*}$ 		  & \cite{hart1968} 		& Informed	& $O(b^{d})$        \\ \hline
  IDA$^{*}$ 	  & \cite{korf1985} 		& Informed	& $O(b^{d})$        \\ \hline
  RBFS 	  	  & \cite{korf1991} 		& Informed	& $O(b^{d})$	     \\ \hline
\end{tabular}}
\end{center}
   \caption{Tree Search Algorithm Comparison ($b$ - branching factor, $d$ - depth of a solution, $m$ - maximum depth). \label{table:treeSearchAlgorithmComparison}}
\end{table}

\subsection{Quantum Search \label{sec:quantumSearch}}

Grover's algorithm performs a generic search for a solution using the principles of quantum computation and mechanics \cite{grover1996} \cite{grover1998a}. Suppose we wish to search through a problem's search space of dimension $N$. Also, consider that we are also capable of efficiently perceiving a solution to our problem. This is similar to the NP class of problems whose solutions are verifiable in polynomial time $O(n^{k})$ for some constant $k$, where $n$ is the size of the input to the problem \cite{edmonds1965}. Grover's search algorithm employs quantum superposition and reversible computation in order to query many elements of the search space simultaneously. 

Grover's algorithm was later experimentally demonstrated in \cite{chuang1998}. The algorithm provides a polynomial speed-up when compared with the best-performing classical search algorithms. As previously mentioned, any such classical algorithm requires $O(N)$ time in order to search $N$ elements. Grover's algorithm requires $O(\sqrt{N})$ time, providing a quadratic speedup, which is considerable when $N$ is large. 

\subsubsection{Oracle unitary operator \label{sec:oracleUnitaryOperator}}

In quantum computation mathematical objects known as unitary operators are responsible for the time-evolution of the state of a close quantum system. This is one of the foundational principles behind quantum computation, and is also known as the evolution postulate \cite{kaye2007a}. A black box, also referred to as an oracle \cite{nielsen2000}, representing a unitary operator $U$, is employed in order to indicate, through a reverse of the associated amplitude, which of the values present in an amplitude register corresponds to the searched ones. This process can be performed by adding an additional input bit $c$ to the original $n$-bit input register $x$ and performing a XOR operation. This behaviour is illustrated in Expression \ref{eq:oracle} which employs the ket notation introduced by Paul Dirac \cite{dirac1939} \cite{dirac1981}. Classical reversible circuitry can also be described through such a formulation, albeit without employing the ket notation.
 
 \begin{equation}
	 U : \ket{ x } \ket{ c } \mapsto \ket{ x } \ket{ c \oplus g( x ) }
 \label{eq:oracle}
 \end{equation}
 
Grover's algorithm employs a process of amplitude amplification, known as Grover's iterate, in order to amplify the amplitudes of the solutions and in the process diminish those of the non-solutions. This process is performed by setting the control register $c$ to a specified eigenvector, which, when combined with Grover's iterate can be mathematically proven to perform an inversion about the mean of the amplitudes \cite{kaye2007a}. As a direct result of Grover's iterate, the probability of an answer bearing state increases. However, the amplitude of the solution value is amplified only in a linear way.  If the function $f$ is provided as a black box, then $O(\sqrt{N})$ applications of the black box are necessary in order to solve the search problem with high probability for any input \cite{nielsen2000}. \footnote{A number of improvements have been purposed since Grover's original work \cite{grover2002} \cite{grover2005}. These improvements essentially targeted reduced time complexity bounds for non-query operations and overall robustness. For a number of several novel search related applications please refer to \cite{grover1998a} \cite{grover1998b} \cite{grover1999}.}

Traditionally, classical computation is seen as an irreversible process, a direct consequence from the use of many-to-one binary gates. A logical gate is a function $f: \{0,1\}^{k} \rightarrow \{0,1\}^{l}$ from some fixed number $k$ of input bits to some fixed number $l$ of outputs bits \cite{mano2002}. A computation is said to be reversible if given the outputs we can recover the inputs \cite{toffoli1980a} \cite{toffoli1980b}. Mathematically, a reversible computation corresponds to the concept of a bijective function. It turns out that there is a general mechanism for converting irreversible computations into reversible ones. Each irreversible gate can be made reversible by adding some additional input and output wires \cite{kaye2007a}. This conversion introduces a certain number of inputs and outputs to each irreversible gate. It is this additional information that provides for reversible computation.

Intuitively, it should come as no surprise that an irreversible circuit can be made reversible by substituting each irreversible gate by an equivalent reversible gate \cite{toffoli1980a}. By substituting each element of the circuit with its respective inverse we are able to perform the inverse operation of the original circuit. In practice, this means that if we run the reversed circuit with an output, we will obtain the originating input bit register.

Emil Post's research into the principals of mathematical logic and description of the complete sets of truth functions \cite{post1941} implied that all functions $f: \{ 0, 1 \}^{k} \rightarrow \{ 0, 1\}^{l}$ could be computed by binary circuits employing logical gates $\neg , \lor$ and $\land$. Since it is always possible to build a reversible version of an irreversible circuit, reversible computation can also compute all functions $f: \{ 0, 1 \}^{k} \rightarrow \{ 0, 1\}^{l}$.

\subsubsection{Quantum superpositions \label{sec:quantumSuperpositions}}

In order to gain a ``quantum advantage'', Grover assembles a quantum superposition containing all possible values that should be presented as input to the unitary operator. At its core, the superposition principle simply conveys the notion that multiple quantum states exist at the same time. Let $\ket{ \psi }$ denote the $n$-bit input register presented to unitary operator $U$, then $\ket{\psi}$ takes the form illustrated in Expression \ref{eq:superpositionInputRegisterV1}.

\begin{equation}
\ket{ \psi } = \frac{1}{\sqrt{2^{n}}} \sum_{ x = 0}^{2^{n} - 1 } \ket{ x }
\label{eq:superpositionInputRegisterV1}
\end{equation}

Since unitary operators obey linearity principles we are now in a position to apply unitary operator $U$ to the superposition register. In practice this process means that all values present in the superposition register are processed simultaneously. This operation is illustrated in Expression \ref{eq:unitaryOperatorAppliedToSuperpositionInputRegister}. Additionally, unitary operators as well as input registers can be described by matrices. Accordingly, in light of Expression \ref{eq:unitaryOperatorAppliedToSuperpositionInputRegister}, applying unitary operator $U$ to input register $\ket{\psi}$, can be understood as performing matrix multiplication $U \ket{ \psi }$.

\begin{equation}
U \ket{ \psi } = \frac{1}{\sqrt{2^{n}}} \sum_{ x = 0}^{2^{n} - 1 } U \ket{ x }
\label{eq:unitaryOperatorAppliedToSuperpositionInputRegister}
\end{equation}

\subsection{Hybrid system \label{sec:hybridSystem}}

As previously mentioned, Grover's algorithm performs a generic search by employing a unitary operator $U$. It was designed with the purpose of searching an unstructured collection of registers. I.e. the algorithm's purpose can be understood as looking for binary strings within a collection. Albeit, since binary strings can be used to encode mathematical abstractions, can we build upon this behaviour in order to develop a hybrid system performing classical hierarchical search using Grover's algorithm? This hybrid approach would combine traditional tree search mechanisms with the efficiency gains promised by Grover's algorithm.

Assume that the unitary operator $U$ employed during Grover's iterate is developed in such a way as to recognize potential solutions. Intuitively, it is fairly easy to see that $U$ will need to contemplate the sequence of steps taken in each path of a tree in order to determine if it leads to a solution. However, it needs to do so by restricting itself to a binary representation. This coding strategy serves two direct purposes, namely (i) provide a numerical basis on which the development of $U$ can be built-upon; (ii) being able to translate Grover's output into a set of actions leading to a solution. These actions are to be interpreted as the set of decisions executed at each step of the tree search, leading from the root node to a goal state.

\begin{figure}[h]%[tp]
	\centering
	\includegraphics[width=8cm]{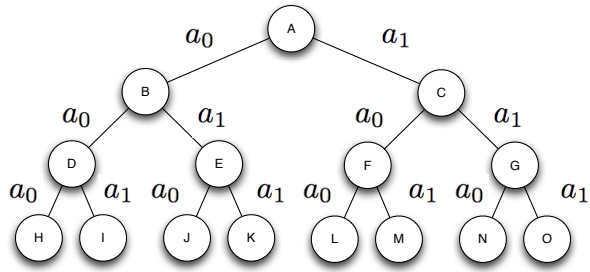}
	\caption{A search tree with a constant branching factor $b = 2$, depth $d = 3$ for a total of $b^d = 8$ leaf nodes\label{fig:constantBranchingFactor}.}
\end{figure}

In order to formally introduce the coding mechanism lets start by considering the binary tree presented in Figure \ref{fig:constantBranchingFactor}. The illustrated binary tree has a root node $A$. Also, has it is possible to see each layer of depth $d$ provides an additional $2^{d}$ nodes to the tree. At each node it is always possible to apply two actions, i.e. we have a branching factor $b = 2$. Each possible path leading to a leaf node can thus be perceived as a concatenation of the binary strings encoding the associated set of performed actions. Using this approach we are able to build a superposition $\ket{ \psi }$ containing all the possible paths. Superposition $\ket{ \psi }$ and a unitary operator $U$ can then be employed by Grover's algorithm to determine the paths leading to solutions.

\subsection{Objectives and Problems \label{sec:objectivesAndProblems}}

As previously mentioned, unitary operators can be derived by transforming irreversible circuits into reversible ones. Accordingly, it is always possible to build a unitary operator $U$ which checks if a set of actions produces a solution. In this work we will not be concerned with the actual implementation details of $U$ but rather on how such a problem could potentially be approached. Additionally, we will also be interested in determining how such an hybrid system would be affected by traditional search concepts. For instance, what are the ramifications of a variable such as the branching factor? Would the system require the use of a constant branching factor? Clearly, this is not always the case for the complete set of problems that can potentially be addressed by search algorithms. When considering a non-constant branching factor, what would be the associated impacts in overall system performance? Additionally, traditional search strategies typically employ some kind of information to determine which states are more promising than others when trying to reach a goal state. Can an heuristic bounded procedure be incorporated into such an approach? If so, do we stand to gain any significant advantage?

These and other questions will be addressed in the remaining sections of this work. In Section \ref{sec:branchingFactorRamifications} we will focus on assessing how such an hybrid system would perform from a branching factor perspective. In Section \ref{sec:heuristicPerspective} we devote our efforts to researching on the impacts of incorporating heuristic concepts into the hybrid proposal. Section \ref{sec:finalConsiderations} presents and discusses the parallels, alongside the differences, between our proposal and another well-known kind of graph inspection tool, respectively the quantum random walk. We present the conclusions of this work in Section \ref{sec:conclusions}. With these sections we hope to lend some intuition into the advantages, and disadvantages, of quantum computation and a possible model for hybrid hierarchical quantum search. We will strive for presenting a accessible mathematical analysis of our hybrid approach which takes into account the referred objectives.

%%%%%%%%%%%%%%%%%%%%%%%%%%%%%%%%%%%%%%%%%%%%%%%%%%%%%%%%%%%%%%%%%%%%
% 							ATTENTION : RETHINK THIS TITLE										      %
%%%%%%%%%%%%%%%%%%%%%%%%%%%%%%%%%%%%%%%%%%%%%%%%%%%%%%%%%%%%%%%%%%%%

\section{Branching factor ramifications \label{sec:branchingFactorRamifications}}

In theoretical computer science one possible way to measure a problem's complexity consists in assessing how long a given algorithm takes to find a solution. However, time performance is dependent on a multitude of hardware related factors. Accordingly, it is often more suitable to take appropriate steps to determine the total number of items that are to be evaluated. In the case of a classical tree search this equates to the number of nodes to take into account.  From a classical tree search perspective complexity is expressed in terms of $b$, the branching factor or maximum number of successors of any node; $d$, the depth of the shallowest goal node; $m$, the maximum length of any path in the state space \cite{russell2003}. Section \ref{sec:constantBranchingFactor} focuses on the aspects surrounding a constant branching factor. The requirements for a non-constant branching factor are presented in Section \ref{sec:nonConstantBranchingFactor}. The impact of using a non-constant branching factor is presented in Section \ref{sec:mathematicalAnalysis}.

\subsection{Constant branching factor \label{sec:constantBranchingFactor}}

As previously stated, our proposal for a hybrid quantum tree search system only relied on a constant branching factor $b$. This was mostly due to simplification reasons. However, a constant branching factor requirement is not feasible when considering potential applications of search algorithms. Since, at its essence, our system can be perceived as evaluating a superposition of all possible paths up to a depth level $d$, it is pivotal to determine the impact of a non-constant branching factor in our approach. Lets proceed by examining a couple of examples in order to have a clear understanding of the search process. 

Figure \ref{fig:constantBranchingFactor} illustrates a search tree where at any given node it is always possible to apply two actions (respectively labeled as $a_{0}$ and $a_{1}$), i.e. the branching factor for this search tree is $2$. Recall from Section  \ref{sec:hybridSystem} that we need to encode these actions in a binary fashion. In order to do so we need to determine how many bits $n$ do we require. For an even number of actions we can simply calculate the base-$2$ logarithm. However, we need to take into account that an odd number of actions might be required. In this case we need to map the value into the next largest integer, which can be done through the ceiling function, i.e. $n = \left \lceil log_{2}{|a|} \right \rceil$ bits, where $|a|$ denotes the cardinality of the action set. Notice that the complete range of values allowed with $n$ bits might not be used, i.e. $|a| < 2^{n}$. From our point of view, it is the unitary operator's responsibility to validate whether a binary string is an admissible action.

In the case of the search tree illustrated in Figure \ref{fig:constantBranchingFactor} which possesses a branching factor $b = 2$ actions we need $\lceil log_{2}{2} \rceil = 1$ bit. Accordingly, let value $0$ denote $a_{0}$ and value $1$ represent $a_{1}$. The binary strings encoding the paths leading to each leaf nodes of the search tree illustrated in Figure \ref{fig:constantBranchingFactor} are presented in Table \ref{table:binaryEncondingsConstantBranchingFactor}.

\begin{table}
\centering
	  \begin{tabular}{| c | c c c |} \hline
	   %\text{Path to node }& $p_{1}$ & $p_{2}$ & $p_{3}$  \\ \hline \hline
	   \text{Path to node } & \text{Action at level 1} & \text{Action at level 2}  & \text{Action at level 3} \\ \hline \hline
	   
	  \text{H} & $0$ & $0$ & $0$ \\ \hline
	  
	  \rowcolor[gray]{.8}
	  \text{I} & $0$ & $0$ & $1$  \\ \hline	  
	  
	  \text{J} & $0$ & $1$ & $0$ \\ \hline
	  
	  \rowcolor[gray]{.8}
	  \text{K} & $0$ & $1$ & $1$ \\ \hline
	  	   
	  \text{L} & $1$ & $0$ & $0$ \\ \hline
	  
	  \rowcolor[gray]{.8}
	  \text{M} & $1$ & $0$ & $1$ \\ \hline
	  
	  \text{N} & $1$ & $1$ & $0$ \\ \hline
	  
	  \rowcolor[gray]{.8}
	  \text{O} & $1$ & $1$ & $1$ \\ \hline
		  
	  \end{tabular}
   \caption{Binary encoding for each possible path of the search tree illustrated in Figure \ref{fig:constantBranchingFactor}\label{table:binaryEncondingsConstantBranchingFactor}}
\end{table}

Suppose we wish to perform a search up to depth level $d$. We can easily build a string of $d$ elements, one for each possible depth, with each element requiring a binary representation using $n$ bits. In total, our binary string will employ $n \times d$ bits. 
We are also able to construct a quantum superposition $\ket{ \psi }$ encompassing all the actions to be applied up to depth level $d$ as illustrated by Expression \ref{eq:superpositionInputRegisterV2}. This superposition $\ket{ \psi }$ can then be employed alongside our unitary operator $U$ and Grover's algorithm. 

\begin{equation}
\ket{ \psi } = \frac{1}{\sqrt{2^{n \times d}}}\sum_{ x = 0}^{2^{n \times d} - 1 } \ket{ x }
\label{eq:superpositionInputRegisterV2}
\end{equation}

\subsection{Non-constant branching factor \label{sec:nonConstantBranchingFactor}}

\begin{figure}[h]%[tp]
	\centering
	\includegraphics[width=8cm]{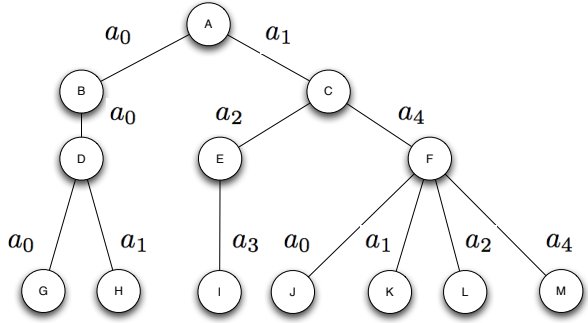}
	\caption{A search tree with a maximum branching factor $b_{max} = 5$ and an average branching factor $b_{avg} = \frac{2 + 1 + 2 + 2 + 1 + 4}{6} = 2$. \label{fig:nonConstantBranchingFactor}}
\end{figure}

Now consider the tree presented in Figure \ref{fig:nonConstantBranchingFactor} with an action set $a = \{a_{0}, a_{1}, a_{2}, a_{3},a_{4}\}$. The first thing one notices is that we no longer have a constant branching factor at each node. In fact for this particular case it is convenient to distinguish between two types of branching factor, namely, the theoretical maximum branching factor $b_{max} = |a| = 5$, and the average branching factor of each node $b_{avg} = 2$, not including the leafs. In order to encode each of the possible actions we require $n = \left \lceil \log_{2}{|a|} \right \rceil = 3$ bits.  Let the encodings of each action be those presented in Table \ref{table:binaryEncondingsProductionNonConstantBranchingFactor}. Accordingly, the binary strings leading to each leaf node are illustrated in Table \ref{table:binaryEncondingsNonConstantBranchingFactor}.

%%%%%%%%%%%%%%%%%%%%%%%%%%%%%%%%%%%%%%%%%%%%%%%%%%%%%%%%%%%%%%%%%%%%
% 				GO THROUGH THE DOCUMENT AND SUBSTITUTE PRODUCTIONS FOR ACTIONS				     %
%%%%%%%%%%%%%%%%%%%%%%%%%%%%%%%%%%%%%%%%%%%%%%%%%%%%%%%%%%%%%%%%%%%%
\begin{table}
\centering
	  \begin{tabular}{| c | c c c |} \hline
	 action	    & $b_{0}$ & $b_{1}$ & $b_{2}$ \\ \hline
	 $a_{0}$               &       $0$ &       $0$ &    $0$ \\ \hline
	 $a_{1}$               &       $0$ &       $0$ &    $1$ \\ \hline
	 $a_{2}$               &       $0$ &       $1$ &    $0$ \\ \hline
	 $a_{3}$               &       $0$ &       $1$ &    $1$ \\ \hline
	 $a_{4}$		    &       $1$ &       $0$ &    $0$ \\ \hline \hline
	 
	 \rowcolor[gray]{.8}
	 undefined            &       $1$ &       $0$ &    $1$ \\ \hline
	 
	 \rowcolor[gray]{.8}
	 undefined            &       $1$ &       $1$ &    $0$ \\ \hline
	 
	 \rowcolor[gray]{.8}
	 undefined            &       $1$ &       $1$ &    $1$ \\ \hline
	 
	  \end{tabular}
   \caption{Binary encoding for each possible path of the search tree illustrated in Figure \ref{fig:nonConstantBranchingFactor}\label{table:binaryEncondingsProductionNonConstantBranchingFactor}}
\end{table}

\begin{table}
\centering
	  \begin{tabular}{| c | c c c |} \hline
	   %\text{Path to node }& $p_{1}$ & $p_{2}$ & $p_{3}$  \\ \hline \hline
	   \text{Path to node } & \text{Action at level 1} & \text{Action at level 2}  & \text{Action at level 3} \\ \hline \hline
	  
	  \text{G} & $000$ & $000$ & $000$ \\ \hline
	  
	  \rowcolor[gray]{.8} 
	  \text{H} & $000$ & $000$ & $001$ \\ \hline
	  
	  \text{I}  & $001$ & $010$ & $011$  \\ \hline	  
	  
	  \rowcolor[gray]{.8}
	  \text{J} & $001$ & $100$ & $000$ \\ \hline
	  
	  \text{K} & $001$ & $100$ & $001$ \\ \hline
	  	   
	  \rowcolor[gray]{.8}	   
	  \text{L} & $001$ & $100$ & $010$ \\ \hline
	  
	  \text{M} & $001$ & $100$ & $100$ \\ \hline		  
	  \end{tabular}
   \caption{Binary encoding for each possible path of the search tree illustrated in Figure \ref{fig:constantBranchingFactor}\label{table:binaryEncondingsNonConstantBranchingFactor}}
\end{table}

In order to employ Grover's algorithm we would need to build a quantum superposition state $\ket{ \psi^{\prime}}$ similar to that presented in Expression \ref{eq:superpositionInputRegisterV2}. Again, our superposition would consist of those states with $d$ elements, each of which with $n$ bits. However, there is a crucial difference between both tree searches. Consider the case illustrated in Figure \ref{fig:nonConstantBranchingFactor} where a search to depth level $d = 3$ alongside $b_{max} = 5$ is performed. Superposition $\ket{ \psi^{\prime}}$ would contain all the quantum states belonging to the range $[0, 2^{d \times n} - 1] = [0, 2^{9} - 1] = [0, 511 ]$. I.e. we would be able encode 512 possible paths when in reality we would only need to encode the states presented in Table \ref{table:binaryEncondingsNonConstantBranchingFactor} \footnote{ An alternative approach would consist in encoding each possible state, instead of encoding in an binary fashion the sequence of actions. However, this would have a meaningful impact on the complexity and design of unitary operator $U$ since each admissible input string would have to be mapped onto a predefined  sequence of actions.}. In reality, the vast majority of the states present in the superposition would contain inadmissible configurations of actions.

It is important to draw attention to the fact that Grover's algorithm provides a quadratic speedup $O( \sqrt{N} )$ where $N$ is the number of elements present in the superposition. By employing $n = \left \lceil \log_{2}{b_{max}} \right \rceil$ bits, when in practice $n = \left \lceil \log_{2}{b_{avg}} \right \rceil$ bits might have sufficed, we are extending the search space and in the process loosing some of the speedup provided by Grover's algorithm. 

Naturally, the question arises: Considering the above encoding mechanism, and a $b_{avg} < b_{max}$ when does Grover stop providing a speedup over classical approaches?

\subsection{Analysis \label{sec:mathematicalAnalysis}}

So, how can we proceed in order to analyze the problem depicted in the previous section? As with so many other fields it is usually easier to start out with an example and extrapolate from that. Accordingly, lets consider the following scenario: we want to perform a tree search using our hybrid quantum search system up to depth level $10$; the maximum branching factor, $b_{max}$ is $5$, however on average we are only able to perform three actions, i.e. $b_{avg} = 3$. Does our approach still provides an advantage over classical search strategies?

In order to answer this question we need to consider the complexities of classical search algorithms. Traditionally, these methods experience some type of exponential growth in the number of leaf nodes that need to be assessed. Since the number of elements to be evaluated is a function of the branching factor $b$ and the depth of the search $d$ the associated complexity is typically of the form $O(b^{d})$. Clearly, in the case of our scenario we need to differentiate between the two branching factors, respectively $b_{max}$ and $b_{avg}$. Accordingly, if we consider $b_{max}$ then a total of $ O( b_{max}^{d} ) = O( 5^{10} ) = 9765625$ nodes might potentially need to be evaluated. On the other hand, by employing $b_{avg}$ a total of $O( b_{avg}^{d} ) = O( 3^{10} ) = 59049$ nodes may be considered. These values differ by a factor of $\approx 165$ which is considerable when in practice it is acceptable to consider $b_{avg}$ as the \textit{de facto} branching factor.

Now lets reflect on the number of times one would need to apply Grover's iterate. In this case there is no distinction to be made between $b_{max}$ and $b_{avg}$ since we would still need to build a quantum superposition state encoding all of the $b_{max}$ actions up to a depth level $d$. As previously stated this would result in a total of $n \times d = 3 \times 10 = 30$ bits being required, where $n = \left \lceil \log_{2}{b_{max}} \right \rceil = \left \lceil \log_{2}{5} \right \rceil = 3$. Accordingly, for the above scenario we would need to apply Grover's iterate a total of $O( \sqrt{N} ) = O( \sqrt{2^{30}} ) = 32768$ times in order to obtain a solution. By comparing the number of states classically evaluated by using $b_{avg}$ against the total number of iterations required by Grover's algorithm we see that both values differ by a factor of $\approx 1.8$. Not surprisingly, although we still obtain a speedup over the classical approach it is severely lessened. Intuitively, it should be clear that this behaviour can be perceived in the following manner: 

\begin{itemize}

	\item As $b_{avg}$ grows closer to $b_{max}$ the number of times to apply Grover's iterate will be optimal relatively to the classical approaches;
	
	\item As $b_{avg}$ grows more distant to $b_{max}$ the number of Grover iterations to apply will be closer to $b_{avg}^{d}$.
	
\end{itemize}

From the above reasoning we are now able to successfully determine the number of times that Grover's iterate should be applied, respectively $|G|$, as illustrated by Expression \ref{eq:numberOfTimesToApplyGrover}.

\begin{align}
|G| 	&= \sqrt{N} \nonumber \\ 
	&= \sqrt{2^{n \times d}} \nonumber \\
	&= \sqrt{2^{\left \lceil log_{2}{b_{max}} \right \rceil \times d}} \nonumber \\
	&= 2^{ \frac{ \left \lceil log_{2}{b_{max}} \right \rceil \times d }{2} }
\label{eq:numberOfTimesToApplyGrover}
\end{align}

Keep in mind that we wish to determine where the threshold lies between the number of elements of a classical search, respectively $b_{avg}^{d}$ and the total number of times to apply Grover's iterate $|G|$. This process can be formulated as presented in Expression \ref{eq:result1} which when solved results in Expression \ref{eq:result2}.

\begin{align}
b_{avg}^{d} 				&= |G| \label{eq:result1} \\
\Leftrightarrow b_{avg}^{d}	&= 2^{ \frac{ \left \lceil log_{2}{b_{max}} \right \rceil }{2}^{d} } \nonumber \\
\Leftrightarrow b_{avg}   		&= 2^{ \frac{ \left \lceil log_{2}{b_{max}} \right \rceil }{2} } \label{eq:result2}
\end{align}

\begin{figure}[h]%[tp]
	\centering
	\includegraphics[width=10cm]{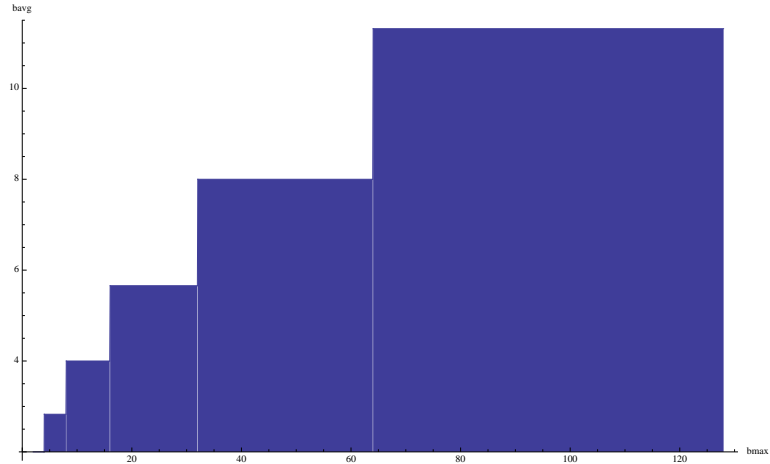}
	\caption{The area plot of $b_{avg} \leq 2^{ \frac{ \left \lceil log_{2}{b_{max}} \right \rceil }{2} } $ for $b_{max} \in [2,128]$. The shaded area indicates those values of $b_{avg}$ that will produce better performance results over our hybrid quantum production system. \label{fig:plotAverageBranchingFactorVsGroverIterations}}
\end{figure}

Accordingly, when $b_{avg} <  2^{ \frac{ \left \lceil log_{2}{b_{max}} \right \rceil }{2} }$ then the total number of nodes evaluated in classical search will be less than the number of times to apply Grover's iterate, i.e. $b_{avg}^{d} < |G|$. Appropriately, when $b_{avg} >  2^{ \frac{ \left \lceil log_{2}{b_{max}} \right \rceil }{2} }$ then our hybrid system will yield a speedup over classical search algorithms. The plot of Expression \ref{eq:result2}, for $b_{max} \in [2,128]$, is presented in Figure \ref{fig:plotAverageBranchingFactorVsGroverIterations}. The shaded area indicates those values of $b_{avg}$ that will produce better performance results classically over the proposed hybrid search. 

Figure \ref{fig:plotAverageBranchingFactorVsGroverIterations} also showcases the characteristic ladder effect of functions employing the ceiling function required by the branching factor binary coding mechanism. Notice that in practice this means that there will always be ranges of values for $b_{max}$ where the number of iterations, $|G|$, will remain the same. Consequently, the average branching factor, $b_{avg}$, for the boundary condition presented in Expression \ref{eq:result1} will also remain constant in those intervals. This happens despite the fact that $b_{max}$ is growing in the associated range.

Lets proceed by elaborating a bit more on each $b_{max}$ and the associated range of $b_{avg}$ values. In order to encode in a binary fashion any $b_{max}$ value we require a total of $n = \lceil log_{2}{b_{max}} \rceil$ bits. The use of the ceiling function effectively forces certain ranges of $b_{max}$ values to require the same number of $n$ bits. As we have seen previously, the number of Grover iterations is a function of the total number of bits required. We can see how this influences $b_{avg}$ by taking Expression \ref{eq:result2} into account. Clearly, for those $b_{max}$ values requiring the same number of bits, the associated $b_{avg}$ values will remain the same. It is only when the required number of bits for $b_{max}$ changes that an impact will be felt regarding $b_{avg}$. Accordingly, we are interested in studying what happens when a transition if performed from, for example, $n$ to $n + 1$ bits. In this case the $b_{avg}$ value grows from $ 2^{\frac{n}{2}}$ to  $2^{\frac{n + 1}{2}}$ which differ by a factor of $\sqrt{2}$. Let $b_{avg}^{n}$ denote the average branching factor $b_{avg}$ associated with an interval requiring $n$ bits. Then, it is possible to express $b_{avg}^{n + 1}$ as a function of $b_{avg}^{n}$. This recurrence behaviour is illustrated in Expression \ref{eq:relationshipBetweenIntervals}.

\begin{equation}
b_{avg}^{n + 1} = \sqrt{2} b_{avg}^{n}
\label{eq:relationshipBetweenIntervals}
\end{equation}

Expression \ref{eq:relationshipBetweenIntervals} can be improved if we allow ourselves to leave behind the use of the ceiling function employed in Expression \ref{eq:result2}. In doing so, we are deliberately abdicating of the ladder effect presented in Figure \ref{fig:plotAverageBranchingFactorVsGroverIterations} and giving the $b_{avg}$ function a quadratic form. This new formulation is presented in Expression \ref{eq:functionWithoutCeiling}.

\begin{equation}
b_{avg} = 2^{ \frac{ log_{2}{ b_{max} } }{2} } = \sqrt{b_{max}}
\label{eq:functionWithoutCeiling}
\end{equation}

\begin{figure}[h]%[tp]
	\centering
	\includegraphics[width=10cm]{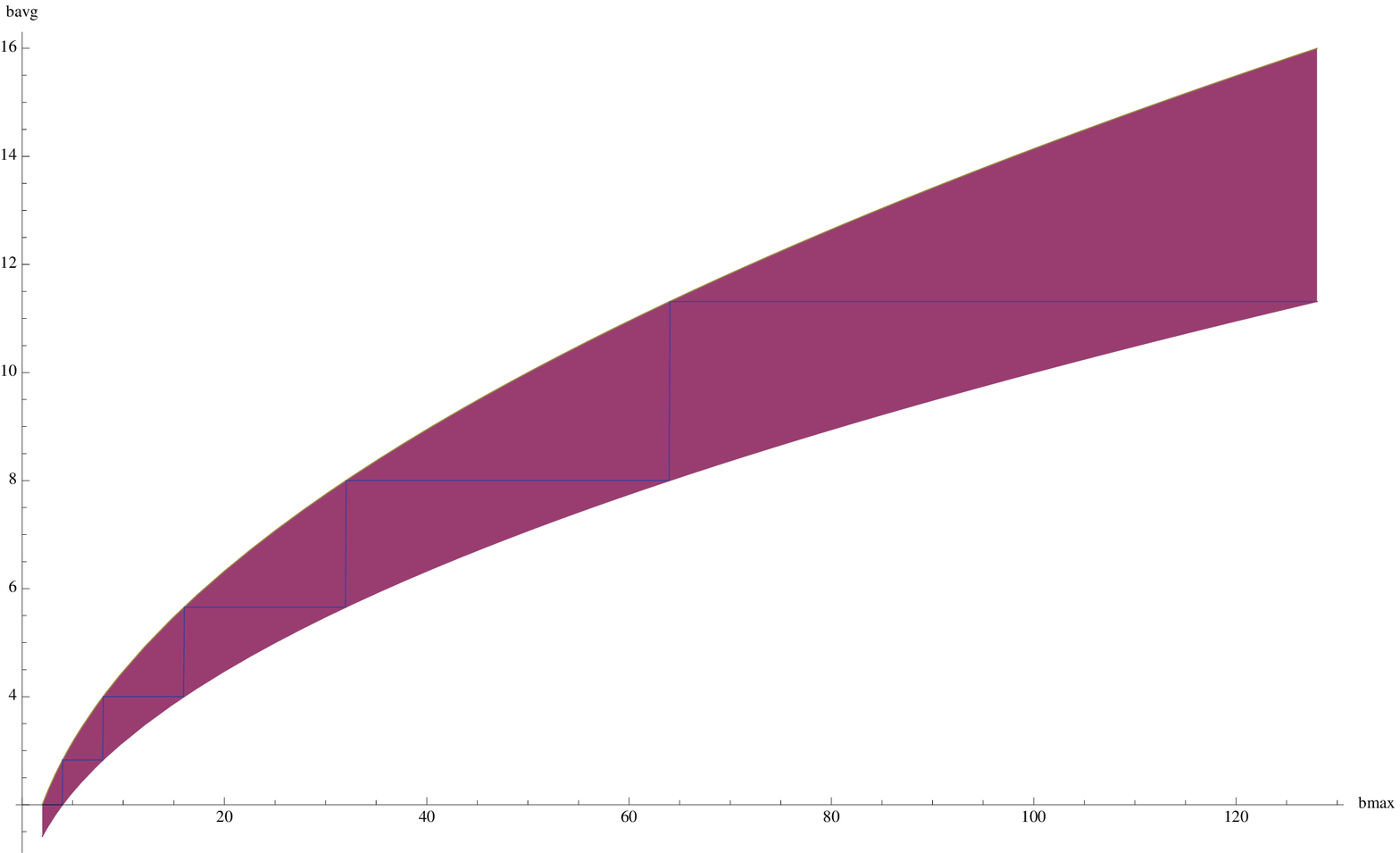}
	\caption{The $\sqrt{2}$-constant growth between $b_{avg}^{\prime}$ and $b_{avg}$ superimposed on the original $b_{avg} = 2^{ \frac{ \left \lceil log_{2}{b_{max}} \right \rceil }{2} } $ function for $b_{max} \in [2,128]$. \label{fig:plotAverageBranchingFactorVsGroverIterationsConstantGrowth}}
\end{figure}

Since we no longer have a ``range-bounded'' function, a direct mapping of Expression \ref{eq:relationshipBetweenIntervals} is impossible. Instead, we can simply define a function, $b^{\prime}_{avg}$, depicting the new upper-range values of $b_{avg}$, as illustrated in Expression \ref{eq:relationshipBetweenFunctions}. This $\sqrt{2}$-constant growth factor is depicted in Figure \ref{fig:plotAverageBranchingFactorVsGroverIterationsConstantGrowth}.

\begin{equation}
b^{\prime}_{avg} = \sqrt{2} b_{avg} = \sqrt{2  b_{max}}
\label{eq:relationshipBetweenFunctions}
\end{equation}

%%%%%%%%%%%%%%%%%%%%%%%%%%%%%%%%%%%%%%%%%%%%%%%%%%%%%%%%%%%%%%%%%%%%
% 							ATTENTION : RETHINK THIS TITLE										      %
%%%%%%%%%%%%%%%%%%%%%%%%%%%%%%%%%%%%%%%%%%%%%%%%%%%%%%%%%%%%%%%%%%%%

\section{Heuristic Perspective \label{sec:heuristicPerspective}}

Determining which production rule should be applied to the working memory is an important part of the control strategy of production system theory. In a great deal of occasions an artificial intelligence application does not possess the adequate level of knowledge allowing for full differentiation amongst the production rules \cite{nilsson1982}. 

At any given point in time during the search process it would be useful to somehow know which production might produce a state which is closer to a goal state. Intuitively, we may describe this process as trying to determine the quality of a path of actions with an optimal solution having the lowest path cost among all solutions \cite{russell2003}. A key component of these systems, and many other algorithms in artificial intelligence, consists in an heuristic function $h(n)$ responsible for presenting an estimate of the distance that a given state $n$ is relatively to a goal state. Function $h(n)$ is typically employed alongside a function $g(n)$ which reflects the search cost incurred to reach state $n$. Traditionally, the conjunction of both these functions is incorporated within a single evaluation function $f( n )$ as illustrated by Expression \ref{eq:evaluationFunction}.

\begin{equation}
f(n) = g(n) + h(n)
\label{eq:evaluationFunction}
\end{equation}

Not surprisingly, function $h(n)$ can have a number of different definitions depending on the specifics of the problem at hand. In a sense, heuristic functions are responsible for providing extra-information about how-well a search is performing. Production systems need not depend on the use of heuristics. Systems which do not take into consideration any kind of auxiliary information are referred to as uninformed strategies. In the case of ``uninformed'' production systems the choice of which rule to apply is performed at random. On the other hand, ``informed'' strategies enable a production system to select an appropriate rule.

Our quantum production system model can be perceived as systematically trying to apply actions in an uninformed fashion. This operation is performed until a goal state is reached. Such behaviour resembles that of a standard blind depth-limited search process. Intuitively, the heuristic concepts incorporated into informed search strategies appear to be an adequate extension idea to our quantum production system. However, it remains to be seen if these concepts can be adequately mapped into our approach and if doing so provides an advantage. The remainder of this section is organized as follows: Section \ref{sec:theQuantumHeuristic}) considers how to incorporate the use of an heuristic the unitary operator; Section \ref{sec:interpretation}) provides an analysis of the quantum computation procedure incorporating the use of an heuristic. Section \ref{sec:extentingTheQuantumHeuristic}) builds on these results to present an extended quantum heuristic mechanism.

\subsection{The Quantum Heuristic \label{sec:theQuantumHeuristic}}

As previously mentioned, at its core, our system can be perceived as evaluating a superposition of all possible paths. Expression \ref{eq:superpositionInputRegisterV2} illustrated this perspective, where a quantum superposition state is composed by an amplitude value and a multitude of sub-states, i.e. the computational basis. From a quantum mechanics perspective it is important to reinforce the idea that the mechanisms allowing quantum states of a superposition to communicate with the other states are complex and restricted in what they achieve. Those algorithms that are able to provide a meaningful speedup do so by employing, and determining, some type of global property. E.g. Grover's algorithm is able to determine the mean amplitude $A$ of a quantum superposition and perform an inversion about the mean \cite{grover1996}. Also Shor's algorithm for factoring numbers is able to efficiently determine the period of periodic quantum states \cite{shor1994}. Accordingly, there is no clear method to communicate between states.

This fact immediately poses a problem: traditionally, the use of heuristics is employed to choose among possible tree paths, ideally producing an optimal sequence of actions. However, with our system, we are unable to quantum mechanically perform such a comparison. Ergo, is there any way to incorporate any heuristic concepts into our hybrid approach? 

From a simplified point of view, an heuristic function outputs a value estimating the distance to a goal. We can therefore opt to consider only those states whose $f$ value is below a certain threshold $T$. Notice that in doing so we are deliberately ignoring the fact that these same states may later on provide an optimal solution path. In order to for the search process to incorporate the heuristic function we need to reflect it upon the unitary operator's design. Recall that a unitary operator $U$ in order to be employed by Grover's algorithm is only required to flip the amplitudes of the solutions states. Expression \ref{eq:unitaryOperatorWithHeuristic} reflects both of the previous requirements, where $\ket{ n }$ represents the current node being processed and $a_{i}$ an action taken at depth $i$.

\begin{equation}
	U \ket{ b } \ket{ a_{1} a_{2} \cdots a_{d} } = 
	\left\{
		\begin{array}{lll}
		-&\ket{ b } \ket{ a_{1} a_{2} \cdots a_{d} } & \text{if} f(b, a_{1}, a_{2}, \cdots, a_{d}) \leq T \\
		
		&\ket{ b } \ket{ a_{1} a_{2} \cdots a_{d}}  & \text{otherwise}
		\end{array} \right.
	\label{eq:unitaryOperatorWithHeuristic}
\end{equation}

In a similar way as to deciding which of two possible heuristics to employ, the matter of deciding the threshold value $T$ could be left to statistical studies, or even informed intuition based on hands-on experience \cite{nilsson1982}. Take notice that in no way did we circumvent the problem of comparing multiple paths. In conclusion, we are able to incorporate some of the concepts surrounding the use of heuristics, but not all of them.

\subsection{Interpretation \label{sec:interpretation}}

What happens when we employ a unitary operator $U$ with the form presented in Expression \ref{eq:unitaryOperatorWithHeuristic} alongside Grover's algorithm? I.e. do we stand to gain anything by incorporating heuristic concepts into our hybrid search system? Section \ref{sec:decomposingTheSuperpositionState} illustrates how superposition $\ket{ \psi }$ can be decomposed into two components. Section \ref{eq:heuristicImpact}) provides a graphical illustration of the impacts of performing search up to different depth levels.

\subsubsection{Decomposing the superposition state \label{sec:decomposingTheSuperpositionState}}

Answering this question requires extending our analysis of what happens when unitary operator $U$ is applied to superposition $\ket{ \psi }$, i.e. $ U \ket{ \psi }$. Suppose superposition $\ket{ \psi }$ takes the form illustrated in Expression \ref{eq:standardQuantumSuperposition}, where $n$ represents the number of bits.

\begin{equation}
	\ket{ \psi } = \sqrt{ \frac{1}{2^{n}} } \sum_{x = 0}^{2^{n} - 1} \ket{x}
	\label{eq:standardQuantumSuperposition}
\end{equation}

Before advancing any further, notice that in an uniform superposition, associated with each state $\ket{x}$ there is an amplitude $\alpha \in \mathbb{C}$, which in this case is $\sqrt{ \frac{1}{2^{n}}}$ for all states. Let $\alpha_{i}$ denote the amplitude associated with state $\ket{i}$. Quantum computation requires the norm of amplitudes to be unit-length, i.e. $\sum_{x = 0}^{2^{n} - 1} |\alpha_{x}|^{2} = 1$, at all times. Not surprisingly any unitary operator must also preserve the norm.

Now note that we can decompose our initial quantum superposition $ \ket{ \psi } $ into two parts \cite{kaye2007a}. One part will contain all those states that are solutions, which we will respectively label as set $X_{good}$. Another part will include the non-solutions states, respectively labeled as set $X_{bad}$. Also, assume that the problem we are interested in solving contains $k$ solutions. This implies that $|X_{good}| = k$ and  $|X_{bad}| = 2^{n} - k$. Accordingly, an uniform superposition of the states belonging to $X_{good}$ would set an equal amplitude among its $k$ elements, i.e. $\sqrt{ \frac{1}{k} }$. By the same reasoning, an uniform superposition of the states in $X_{bad}$ would impose an amplitude values of $\sqrt{ \frac{1}{2^{n} - k} }$. Accordingly, we can define the superposition states $\ket{ \psi_{good} }$ and $\ket{ \psi_{bad} }$, respectively presented in Expression \ref{eq:psiGood} and Expression \ref{eq:psiBad}.

\begin{equation}
	\ket{ \psi_{good} } = \sqrt{ \frac{1}{k} } \sum_{x \in X_{good}} \ket{x}
	\label{eq:psiGood}
\end{equation}

\begin{equation}
	\ket{ \psi_{bad} } =  \sqrt{ \frac{1}{2^{n} - k} } \sum_{x \in X_{bad}} \ket{x}
	\label{eq:psiBad}
\end{equation}

We are now able to express superposition $\ket{ \psi }$ in terms of the subspaces  $\ket{ \psi_{good} }$ and $\ket{ \psi_{bad} }$. This process is illustrated in Expression \ref{eq:superpositionReformulated} which is indispensable to the rest of our analysis.

\begin{equation}
	\ket{ \psi } = \sqrt{ \frac{k}{2^{n}} } \ket{ \psi_{good} } + \sqrt{ \frac{2^{n} - k}{2^{n}}} \ket{ \psi_{bad}}
	\label{eq:superpositionReformulated}
\end{equation}

The requirement that the norm must be preserved at all times induces a probabilistic behaviour. Consider state $\ket{ \psi }$ whose norm is $\left | \sqrt{ \frac{k}{2^{n}} } \right |^{2} + \left | \sqrt{ \frac{2^{n} - k}{2^{n}}} \right  |^{2}  = 1$. I.e. we have a sum of values which sum up to $1$ similarly to a probability distribution. Accordingly, the probability of obtaining state $\ket{ \psi_{good}} = \left |\sqrt{ \frac{k}{2^{n}} } \right |^{2}$ and state $ \ket{ \psi_{bad}} = \left | \sqrt{ \frac{2^{n} - k}{2^{n}}} \right  |^{2}$. Applying Grover's iterate effectively changes the amplitudes, maximizing the probability of obtaining a solution contained in the subspace spawned by $\ket{ \psi_{good} }$. As a result, the amplitude associated with state $\ket{ \psi_{good} }$ will increase. Since the norm of state $\ket{ \psi }$ must be preserved, employing Grover also implies that the amplitude of $\ket{ \psi_{bad} }$ will be decreased.

\subsubsection{Heuristic Impact \label{eq:heuristicImpact}}

In order to continue with our analysis lets assume we possess an admissable heuristic which always eliminates candidate states with each additional level of depth. Although this is a rather optimistic strategy the heuristic's behaviour can be viewed as ideal. We will use this  best case scenario to demonstrate the system's potential performance.

Given a sufficiently high depth level $d$ the heuristic will have to eventually produce the exact number of solutions $k$. Let $k_{d}$ denote the number of solutions at depth level $d$. Then, with such an heuristic we would have $k_{1} < k_{2} < \cdots < k_{d} \leq k$. However, we can never produce fewer than $k$ solutions, since that would violate basic assumptions about the search space of the problem. 

\begin{figure}[h]%[tp]
	\centering
	\includegraphics[width=8cm]{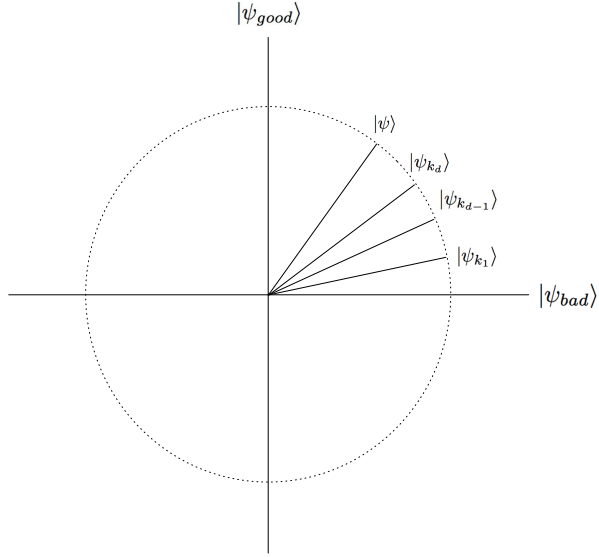}
	\caption{The geometric interpretation of the original state vector $\ket{ \psi }$ alongside different state vectors $\ket{ \psi_{i}}$ reflecting the state attained by performing a search to depth-level $i$. Deeper searches will be closer to the original $\ket{\psi}$ state. All states are unit-length vectors. \label{fig:geometricVisualization}}
\end{figure}

The above process can be visualized geometrically. Notice that our initial superposition $\ket{ \psi }$ containing $k$ solutions can be understood as vector with two components $( \ket{ \psi_{good}}, \ket{ \psi_{bad}} ) = ( \sqrt{ \frac{k}{2^{n}} },  \sqrt{ \frac{2^{n} - k}{2^{n}}})$. Therefore, we are able to map it into a two dimensional plane with axis $ \ket{ \psi_{good}}$ and $\ket{ \psi_{bad}}$. This mapping process is illustrated in Figure \ref{fig:geometricVisualization}. From a quantum computation perspective, we can envisage the use of different states $\ket{ \psi_{k_{d}} }$ reflecting a search up to depth level $d$. State $\ket{ \psi_{k_{d}} }$ is a simple reformulation in terms of $k_{d}$ of state $\ket{\psi}$, as presented in Expression \ref{eq:superpositionReformulatedInTermsOfSolutionsPerDepth}.

\begin{equation}
	\ket{ \psi_{k_{d}} } = \sqrt{ \frac{k_{d}}{2^{n}} } \ket{ \psi_{good} } + \sqrt{ \frac{2^{n} - k_{d}}{2^{n}}} \ket{ \psi_{bad}}
	\label{eq:superpositionReformulatedInTermsOfSolutionsPerDepth}
\end{equation}

Clearly, as the number of solutions $k_{d}$ grows,  to the allowable maximum of $k$, the $\ket{\psi_{good}}$ and $\ket{\psi_{bad}}$ components will tend towards the values of the original state vector $\ket{\psi}$. This behaviour is illustrated in Expression \ref{eq:limitsStudy}.

\begin{equation}
	\lim_{k_{d} \rightarrow k } \left( \sqrt{ \frac{k_{d}}{2^{n}} },  \sqrt{ \frac{2^{n} - k_{d}}{2^{n}}}  \right) = \left( \sqrt{ \frac{k}{2^{n}} },  \sqrt{ \frac{2^{n} - k}{2^{n}}}  \right)
	\label{eq:limitsStudy}
\end{equation}

Not surprisingly, as the depth of the search increases the corresponding state vector $\ket{ \psi_{k_{d}} }$ gets closer to the original $\ket{ \psi }$, as can be perceived from Figure \ref{fig:geometricVisualization}. Additionaly, each state $\ket{ \psi_{k_{d}} }$ can be understood as forming an angle $\theta$ whose tangent is shown in Expression \ref{eq:angleTangent}.

\begin{equation}
	\tan \theta_{k_{d}} = \frac{ \sqrt{ \frac{k_{d}}{2^{n}} } }{  \sqrt{ \frac{2^{n} - k_{d}}{2^{n}}} }   = \sqrt{\frac{k_{d}}{2^{n} - k_{d}}}
	\label{eq:angleTangent}
\end{equation}

Using this approach we can perform a comparison between the angles of different search depth levels. Lets say we wish to compare how the search advanced between depth levels  $d_{1}$ and $d_{2}$. We can define an operator $\Delta \theta_{k_{d_{1}}, k_{d_{2}}}$ with the behaviour defined in Expression \ref{eq:angleOperator}.

\begin{equation}
	\Delta \theta_{k_{d_{1}}, k_{d_{2}}} := \underbrace{\arctan{\sqrt{\frac{k_{d_{2}}}{2^{n} - k_{d_{2}}}}}}_{\theta_{k_{d_{2}}}} - \underbrace{\arctan{\sqrt{\frac{k_{d_{1}}}{2^{n} - k_{d_{1}}}}}}_{\theta_{k_{d_{1}}}}
	\label{eq:angleOperator}
\end{equation}

Operator $\Delta \theta_{k_{d_{1}}, k_{d_{2}}}$ provides a mechanism for determining the operational success of adding $(d_{2} - d_{1})$ extra levels of depth relatively to $d_{1}$. Accordingly, small $\Delta \theta_{k_{d_{1}}, k_{d_{2}}}$ values can be understood as not contributing in a significant manner to changing the system's overall state. Conversely, high $\Delta \theta_{k_{d_{1}}, k_{d_{2}}}$ values reveal that the system's state significantly shifted towards $\ket{\psi}$. 

Ultimately, we can only expect to get as far as state $\ket{\psi}$. From this point on Grover's usual $O(\sqrt{N})$ iterations will still be required in order to perform a rotation of $\ket{ \psi }$ towards $\ket{ \psi_{good}} $, leaving the algorithms overall complexity unchanged. %In conclusion, although the proposed heuristic extension might seem an apparently good addition to our current model it fails to provide any kind of significant advantage.

\subsection{Extending the quantum heuristic \label{sec:extentingTheQuantumHeuristic}}

Can the concepts of the quantum heuristic be extended in such a way as to perform some kind of useful task? In order to answer this question we will assume that the heuristic function employed has some kind of probabilistic distribution. Accordingly, we will start by reviewing some principles surrounding heuristic distributions. We will then show how to incorporate these concepts alongside our hybrid quantum search proposal.

\subsubsection{Heuristic distributions \label{eq:heuristicDistributions}}

Lets assume that the heuristic function employed has a probabilistic distribution, i.e. the probability of each output value is between $0$ and $1$. This behaviour can be written as $ 0 \leq P( X = x ) \leq 1$ where $X$ is a random variable representing an output event, and $x$ is a possible value of $X$. Additionally, a random variable may be either discrete or continuous depending on the values that it can assume. Random variables whose set of possible values belong to $\mathbb{Z}$ are said to be discrete. On the other other hand, continuous variables map events to an uncountable set such as $\mathbb{R}$.  For a discrete random variable $X$, the sum of the set containing all possible probability values is $1$ as illustrated by Expression \ref{eq:discreteRandomVariable}. 

\begin{equation}
\sum_{i = 1}^{n} P( X = x_{i} ) = 1
\label{eq:discreteRandomVariable}
\end{equation}

\begin{figure}[h]%[tp]
  \centering
  	\subfloat[Initial board configuration]	{\label{fig:3x3slidingBlockPuzzle1} \includegraphics[width=0.3\textwidth]{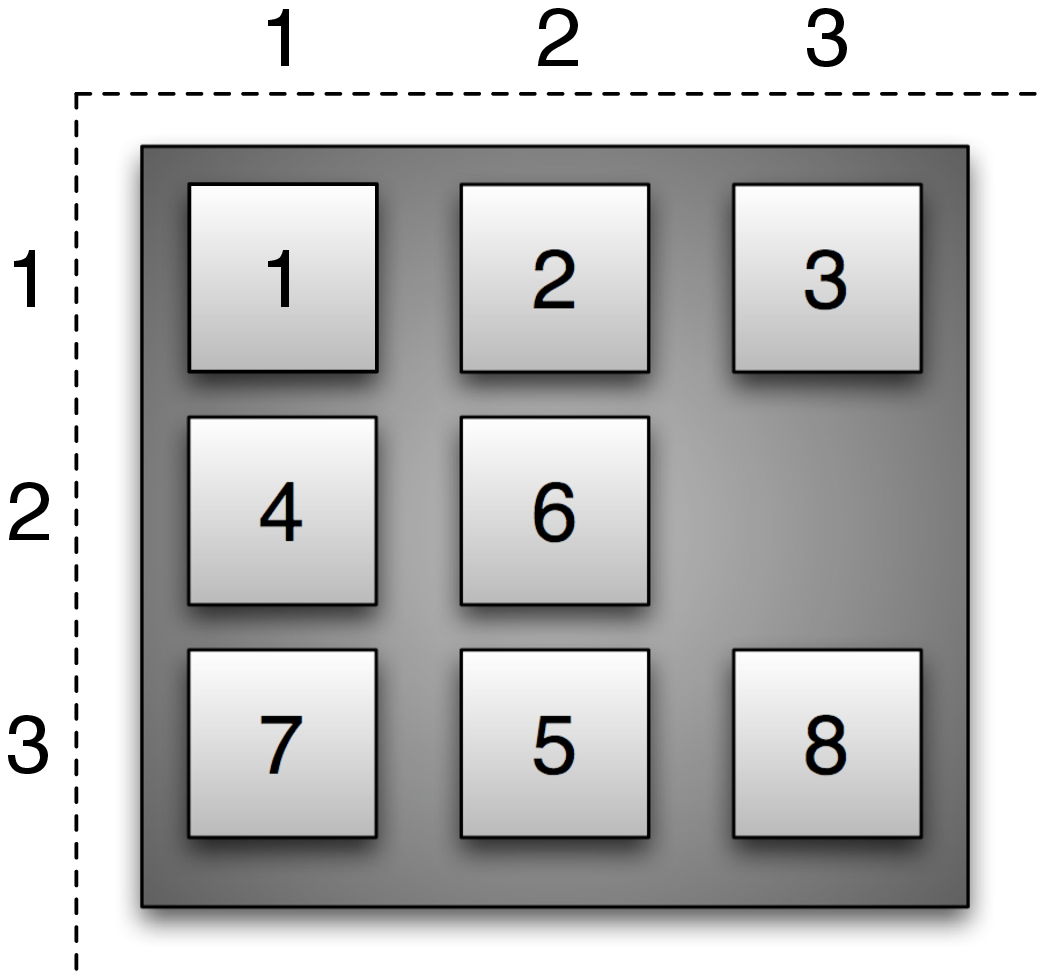}}               
 	\subfloat[End board configuartion]   {\label{fig:3x3slidingBlockPuzzle2} \includegraphics[width=0.3\textwidth]{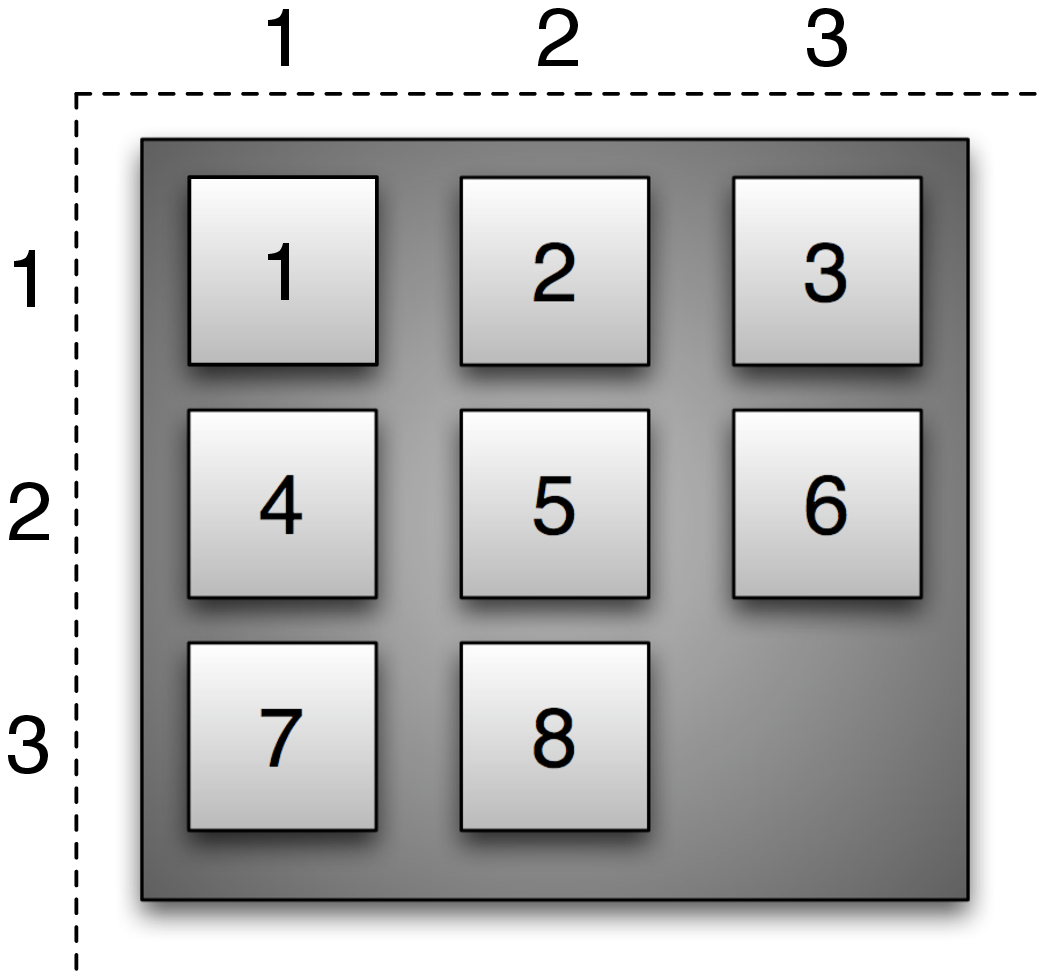}}
  \caption{A sliding block puzzle example with a board of dimension $3 \times 3$, also known as the $8$-puzzle.}
  \label{fig:3x3slidingBlockPuzzle}
\end{figure}

\begin{figure}[!h]
\centering
\includegraphics[width=10cm]{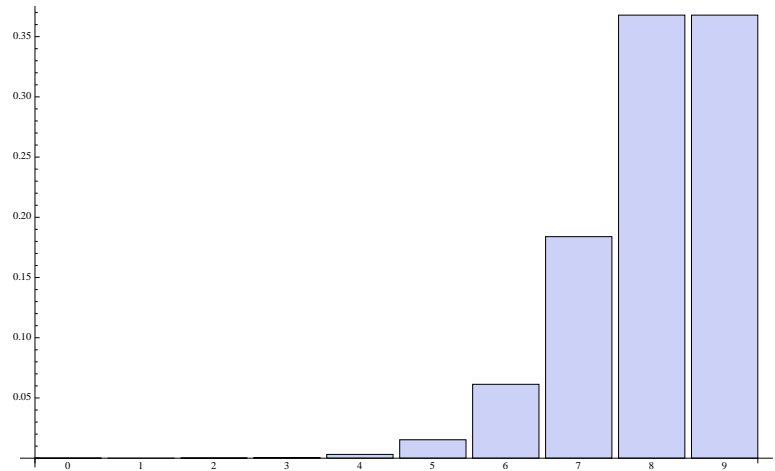}
\caption{Discrete probability distribution for $h_{1}$ when applied to the $8$-puzzle \label{fig:discreteHeuristic8Puzzle}}
\end{figure}

As a concrete example of a discrete heuristic we can consider the sliding block puzzle, also known as the $n$-puzzle, search problem. A sliding block puzzle challenges a player to shift pieces around on a board without lifting them to establish a certain end-configuration, as illustrated in Figure \ref{fig:3x3slidingBlockPuzzle}. This non-lifting property makes finding moves, and the paths opened up by each move important parts of solving sliding block puzzles \cite{hordern1987}. Accordingly, we can define a  heuristic function $h_{1}$ which simply calculates the number of misplaced tiles between a board and a target board configuration. Function $h_{1}$ outputs values belonging to the range $[0, 9]$ and consequently can be classified as discrete. The discrete probability distribution for function $h_{1}$ regarding the $8$-puzzle is presented in Figure \ref{fig:discreteHeuristic8Puzzle}.

If $X$ is a continuous random variable then there exists a nonnegative function $P(X)$, the probability density function of the random variable $X$. The density function $P(X=c)$ is defined as the ratio of the probability that $X$ falls into an interval around $c$, divided by the width of the interval, as the interval width goes to zero \cite{degroot2002}. This process is illustrated in Expression \ref{eq:probabilityDensityFunction}. Additionally, the density function must be nonnegative for all arguments and must obey the behaviour presented in Expression \ref{eq:probabilityDensityFunctionMustHaveIntegralEqualToOne} \cite{russell2003}.

\begin{equation}
P( X = c ) = \lim_{dx \to 0} P( c \leq X \leq c + dx ) / dx
\label{eq:probabilityDensityFunction}
\end{equation}

\begin{equation}
\int_{-\infty}^{+\infty} P(X) dx = 1
\label{eq:probabilityDensityFunctionMustHaveIntegralEqualToOne}
\end{equation}

\begin{figure}[!h]
\centering
\includegraphics[width=10cm]{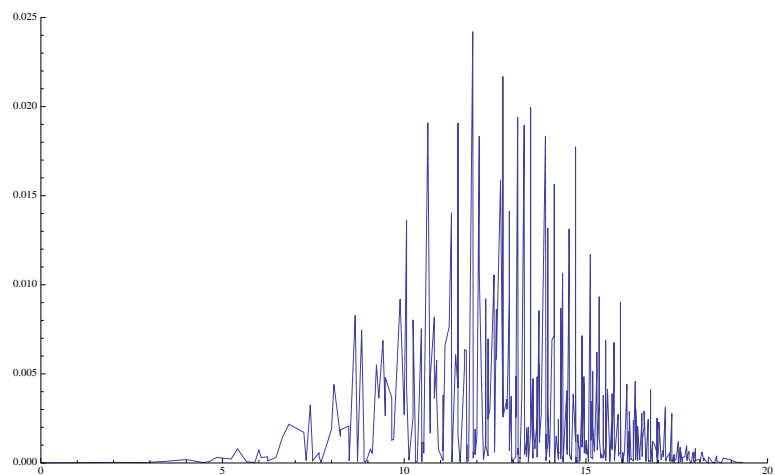}
\caption{Continuous probability density function for $h_{2}$ when applied to the $8$-puzzle \label{fig:continuousHeuristic8Puzzle}}
\end{figure}

As an example of a continuous heuristic function we can again consider the $n$-puzzle search problem. This time we need only to determine an heuristic function $h_{2}$ mapping values into a real codomain. This can be done by employing a metric such as the  euclidean distance. For instance, we can define $h_{2}$ in such a fashion as to calculate the euclidean distance for all the corresponding elements of a board and a target board configuration. I.e., $ h_{2} = \sum_{ \forall{ \text{ tiles}} } d( l_{\text{tile}}, l_{\text{tile}}^{\prime})$, where $d$ is the euclidean distance, $l_{\text{tile}}$ the location of a tile in a board configuration and $l_{\text{tile}}^{\prime}$ the location of the corresponding tile in the target configuration. The probability density function for $h_{2}$ regarding the $8$-puzzle is presented in Figure \ref{fig:continuousHeuristic8Puzzle}.

The probability that a random variable $X$ takes on a value that is less than or equal to $x$ is referred to as the cumulative distributive function $F$ which has the form shown in Expression \ref{eq:cumulativeDistributiveFunction}.

\begin{equation}
F( x ) = P( X \leq x )
\label{eq:cumulativeDistributiveFunction}
\end{equation}

The cumulative distribution function $F(a)$ for a discrete random variable is a simple sum of the values up to element $a$. This behaviour is illustrated in Expression \ref{eq:cumulativeDistributionFunctionDiscreteRandomVariable}. Similarly, a cumulative probability density function $F(a)$ can be defined for a continuous random variable as shown in Expression \ref{eq:cumulativeDistributionFunctionContinuousRandomVariable}.

\begin{equation}
F(a) = \sum_{\text{all } x \leq a} P( X = x )
\label{eq:cumulativeDistributionFunctionDiscreteRandomVariable}
\end{equation}

\begin{equation}
F(a) = \int_{-\infty }^{a} P(x) dx
\label{eq:cumulativeDistributionFunctionContinuousRandomVariable}
\end{equation}

Now suppose we wish to determine when, that is for which values, the cumulative distribution function equals a certain probability $p$, i.e. $F( x ) = p$. In probability theory, this mapping is known as the quantile function of a cumulative distribution function $F(x)$ and is expressed as $F^{-1}(p) = x$. Once more we need to be careful in order to differentiate between discrete and continuous random variables. In the former, there may exist gaps between values in the domain of the cumulative distribution function, accordingly $F^{-1}$ is defined as presented in Expression \ref{eq:quantileFunctionDiscreteRandomVariable} where $inf$ denotes the infimum operator \cite{gilchrist2000}.

\begin{equation}
F^{-1}(p) = inf\{ x \in \mathbb{R} : p \leq F(x) \}
\label{eq:quantileFunctionDiscreteRandomVariable}
\end{equation}

In the case of a continuous random variable obtaining a clear expression for the quantile function is not so trivial since it requires determining the inverse of an integral. Notwithstanding, it is still possible to determine such expressions, for instance, the quantile function of a normal distribution with mean $\mu$ and standard deviation $\sigma$ is illustrated in Expression \ref{eq:quantileFunctionNormalDistribution}, where $erf$ is the Gauss error function \cite{gilchrist2000}.

\begin{equation}
F^{-1}( p, \mu, \sigma^{2} ) = \mu + \sqrt{2} \sigma erf^{-1}( 2p - 1), \quad p \in [0, 1]
\label{eq:quantileFunctionNormalDistribution}
\end{equation}

%%%%%%%%%%%%%%%%%%%%%%%%%%%%%%%%%%%%%%%%%%%%%%%%%%%%%%%%%%%%%%%%%%%%
% ATTENTION: THERE IS A PAPER \cite{kwan1996} THAT ASKS WHETHER NORMAL DISTRIBUTIONS ARE APPROPRIATE %
% TO MODEL CONSTRAINT SATISFACTION PROBLEMS WHICH ARE TYPICALLY SOLVED THROUGH SEARCH                   %
% ALGORITHMS. THE AUTHOR ARRIVES TO THE CONCLUSION THAT IT IS NOT APPROPRIATE TO DO SO, NAMELY:       %
%``We have provided evidence in this paper that the normality assumption should not be taken for granted in CSP research.''    %
%%%%%%%%%%%%%%%%%%%%%%%%%%%%%%%%%%%%%%%%%%%%%%%%%%%%%%%%%%%%%%%%%%%%

\subsubsection{Extended quantum heuristic \label{extendedQuantumHeuristic}}

Naturally, the question arises, how can this process be combined with our hybrid quantum search approach? It turns out that when a fourth of all possible states analyzed by Grover's algorithm are marked as solutions, i.e. $k = \frac{1}{4} . N$ then a single iteration is required in order to obtain with certainty one of the solution states \cite{hirvensalo2004}. Accordingly, we can try to fine tune the behaviour of our unitary operator such that it outputs a solution for a fourth of all cases. This procedure can be performed with the assistance of the quantile function concept introduced in the previous section.

Lets say we have an heuristic function $f: X \to Y$ and are interested in obtaining the states which are closest to a solution. Accordingly, we are interested in marking as a solution those states that produce the smallest values of codomain $Y$. Lets assume that the states which are closer to a goal node are those who tend to happen less times (as exemplified by Figure \ref{fig:discreteHeuristic8Puzzle} and Figure \ref{fig:continuousHeuristic8Puzzle}). From a probabilistic point-of-view it is possible to check if the heuristic value is less than or equal to the quantile function output for a probability of $25\%$. This behaviour is illustrated in Expression \ref{eq:unitaryOperatorWithExtendedHeuristicV1}.

\begin{equation}
	U \ket{ b } \ket{ a_{1} a_{2} \cdots a_{d} } = 
	\left\{
		\begin{array}{lll}
		-&\ket{ b } \ket{ a_{1} a_{2} \cdots a_{d} } & \text{if} f(b, a_{1}, a_{2}, \cdots, a_{d}) \leq F^{-1}(0.25) \\
		
		&\ket{ b } \ket{ a_{1} a_{2} \cdots a_{d}}  & \text{otherwise}
		\end{array} \right.
	\label{eq:unitaryOperatorWithExtendedHeuristicV1}
\end{equation}

Ideally, by using this approach it is possible to obtain a superposition containing one fourth of the ``closest'' states to a goal configuration by applying a single iteration of Grover's algorithm. Additionally, we can further expand on the results of Expression \ref{eq:unitaryOperatorWithExtendedHeuristicV1} in order to contemplate different sections of a probability distribution. For instance, we can choose to obtain in a single iteration of Grover's algorithm the states which lie between heuristic values $(F^{-1}(0.5) - F^{-1}(0.25))$. Similarly, we could also choose to obtain the states belonging to $(F^{-1}(0.75) - F^{-1}(0.5))$ or $(F^{-1}(1) - F^{-1}(0.75))$. More generally, let $a$ and $b$ denote two probabilities values such that $b - a = 0.25$, then it is possible to determine if a given heuristic value belongs to the range $[F^{-1}(a), F^{-1}(b)]$. This strategy for selecting $25\%$ of the search space is presented in the unitary operator shown in Expression \ref{eq:unitaryOperatorWithExtendedHeuristicV2}, 

\begin{equation}
	U \ket{ b } \ket{ a_{1} a_{2} \cdots a_{d} } = 
	\left\{
		\begin{array}{lll}
		-&\ket{ b } \ket{ a_{1} a_{2} \cdots a_{d} } & \text{if} f(b, a_{1}, a_{2}, \cdots, a_{d}) \in [F^{-1}(a), F^{-1}(b)]  \\
		
		&\ket{ b } \ket{ a_{1} a_{2} \cdots a_{d}}  & \text{otherwise}
		\end{array} \right.
	\label{eq:unitaryOperatorWithExtendedHeuristicV2}
\end{equation}

In a certain sense, employing this type of procedure allows for a kind of partial selection of the search space to be performed using a single Grover iteration. The selection is only partial because upon measurement a random collapse amongst the marked states is obtained.

\section{Final considerations \label{sec:finalConsiderations}}

%%%%%%%%%%%%%%%%%%%%%%%%%%%%%%%%%%%%%%%%%%%%%%%%%%%%%%%%%%%%%%%%%%%%
% 			CONSIDER REWRITING THIS CONSIDERATION OF THE QUANTUM RANDOM WALKS       			     %
%%%%%%%%%%%%%%%%%%%%%%%%%%%%%%%%%%%%%%%%%%%%%%%%%%%%%%%%%%%%%%%%%%%%
It is important to mention that some of the graph dynamics considered in this work relate directly to the well-studied quantum random walks on graphs (for an introduction to this research area please refer to \cite{ambainis2003}, \cite{kempe2003b} and \cite{ambainis2004}. Quantum random walks are the quantum equivalents of their classical counterparts (we refer the reader to \cite{hughes1995} \cite{woess2000} for basic facts regarding random walks). Quantum random walks were initially approached in \cite{aharonov1993}, \cite{meyer1996}, \cite{nayak2000} and \cite{ambainis2001} in one-dimensional terms, i.e. walk on a line. The system is described in terms of a position $n$ on the line and a direction $d$, i.e. $\ket{n}\ket{d}$ in the Hilbert space $\mathcal{H} = \mathcal{H}_{n} \otimes \mathcal{H}_{d}$ where $\mathcal{H}_{n}$ is the Hilbert space spanned by the basis vectors encoding the position and $\mathcal{H}_{d}$ the Hilbert space spanned by the vectors of the direction. The direction register, sometimes referred to as the coin space, is initialized to a superposition of the possible direction, in the case of a walk on a line, either left or right, and the position $n$ updated based on the direction of the walk. The choice of which superposition to apply is also a matter investigated.  Surprisingly, if the system is executed for $t$ steps it behaves rather differently than its classical random walks. Specifically, the authors found that the system spreads quadratically faster over the line than its classical equivalent.

Some of the first approaches proposing quantum random walking on graphs can be found in \cite{farhi1997}, \cite{hogg1998}, \cite{aharonov2001} and \cite{childs2002a}. Let $G(V,E)$ represent a $d$-regular graph, and let $\mathcal{H}_{V}$ be the Hilbert space spanned by states $\ket{v}$ where $v \in V$, and $\mathcal{H}_{E}$ be an Hilbert space of dimension $d$ spanned by basis states ${\ket{1}, \cdots, \ket{d}}$. Overall, the system can be described through basis states $\ket{v}\ket{e}$ for all $v \in V$ and $e \in E$. The common approach is to randomly select one of the edges $e$ adjacent to $v$ and the update the current position of the graph, i.e. $U \ket{v} \ket{e} \rightarrow \ket{v^{\prime}} \ket{e}$, if $e$ has edge points $v$ and $v^{\prime}$. The random selection may also be performed using a $d$-dimensional coin space. Perhaps more interesting is the hitting time, i.e. the time it takes to reach a certain vertex $B$ starting from a vertex $A$. In \cite{childs2002a} and \cite{childs2002b} a graph example is presented where a classical random walk would take $\Omega(2^{d})$ steps to reach $B$, where $d$ is the depth of the graph. However, the quantum equivalent walk can reach $B$ in $O(d^{2})$ computational steps, providing for an exponential speedup! Other examples of quantum random walks include how to adapt the models to perform a search \cite{shenvi2003}, \cite{ambainis2004}, \cite{ambainis2005}, and \cite{ambainis2007}

Not surprisingly, due to the hard computational problems being asked, the vast majority of these approaches put a  strong emphasis on actually determining if a node can be reached, and if so how fast. Consequently, questions about the path leading to a pre-specified node have not been properly addressed. For instance, this fact is explicitly pointed out in \cite{childs2002b}, namely: ``Note that although our algorithm ﬁnds the name of the exit, it does not ﬁnd a particular path from entrance to exit''. For many artificial intelligence applications the ability to answer this question is a crucial one as it provides the basis for powerful inference mechanisms capable of knowledge deduction. The ability to obtain the full path open measurement is a key feature of the production system proposed in this work. Additionally, quantum random walks incorporate previous knowledge of a graph in the form of the relationship $G(V,E)$. This kind of knowledge may not always be available from start, e.g. systems where new nodes are generated based on the information accessible to a system at any given point in time.

The quantum random walk approach differs drastically from the model presented in the previous sections which focused on detailing some of the key notions supporting an hybrid quantum search system. Such a system incorporated classical search concepts, expressed through unitary operators, with Grover's quantum search algorithm. From a classical point of view, applying a search procedure can be understood as partitioning the search space into blocks. Each block is then examined in order to determine if a solution is present. However, such an approach did not in any way performed a kind of partition of the ``quantum search space''. Instead, it executed a sequence of actions in order to determine if a goal state was reached. So the question naturally arises: Is there any procedure to perform a hierarchical quantum search based solely on decomposition of the quantum search space?

Grover and Radhakrishnan were some of the first to purpose a possible approach to this problem in \cite{grover2004a}. The main motivation of the work was the following: consider a quantum search space containing a single solution is divided into $l$-blocks of equal size, suppose that we wish to determine in which of the $l$-blocks the solution is, can this process be performed with fewer queries than the original Grover algorithm? In practice, this problem reduces to the one of determining the first $m$ bits of the $n$ bit computational basis containing the solution, with $m \leq n$. The authors proceed by analysing what happens when a variation of Grover's iterate for amplitude amplification is applied to each block. They conclude that it is indeed easier to determine the initial $m$ bits. However, as $m$ grows closer to $n$ the computational gains obtained disappear \cite{grover2004a}. Later, an optimization to this approach was proposed in \cite{korepin2007}, albeit only marginally improving the lower bound on the number of oracle queries.

A number of authors have also focused on extending Grover's original search algorithm. Namely, in \cite{zalka1999a} the quantum search algorithm was shown to be optimal in the sense that it gives the maximal possible probability of finding a solution. Other examples of possible extensions to Grover's original work include returning a solution with certainty (please refer to \cite{zalka1999b}, \cite{brassard2000}, \cite{long2001} and \cite{hu2002}) and assessing the impact of having multiple solutions \cite{boyer1998}. Additionally, it should be mentioned that no quantum black-box search algorithm can solve the search problem by making fewer than $\Omega( \sqrt{N} )$ iterations. This result was first shown in \cite{bennett1997} and later revised on \cite{boyer1998}.

\section{Conclusions \label{sec:conclusions}}

In this work we examined the ramifications of an hybrid quantum search system. We chose to focus on two key aspects: the use of a non-constant branching factor and the adoption of an heuristic point of view. Clearly, both of these cases are not without its flaws. However, we consider the additional insight provided to be valuable. In the case of the non-constant branching factor we were able to verify that deciding whether or not to Grover, or to proceed classically, should take into account the maximum and the average branching factor. Additionally, by evaluating the states that are within a given threshold we are able to incorporate some classical heuristic concepts into our approach. These concepts were further extended with the use of probabilistic distribution functions allowing for a selection mechanism to be obtained. This mechanism enables specific ranges of quantum states to be obtained in an efficient manner.

\section{Acknowledgements}

This work was supported by FCT (INESC-ID multiannual funding) through the PIDDAC Program funds and FCT grant DFRH - SFRH/BD/61846/2009.

%%%%%%%%%%%%%%%%%%%%%%%%%%%%%%%%%%%%%%%%%%%%%%%%%%%%%%%%%%%%
%       									BIBLIOGRAPHY					         		     %
%%%%%%%%%%%%%%%%%%%%%%%%%%%%%%%%%%%%%%%%%%%%%%%%%%%%%%%%%%%%
%\section{References}
%\bibliographystyle{ieeetr}
\bibliographystyle{apalike}
%\bibliography{../../../../../Bibliography/bibliography}

\end{document}